# An Edge Computing Robot Experience for Automatic Elderly Mental Health Care Based on Voice

C. Yvanoff-Frenchin[1], V. Ramos[2], T. Belabed[3], C. Valderrama[4]

[1] ESIEE, France, camille.yvanoff-frechin@edu.esiee.fr

[2] UMons, Belgium, vitor.ramos@umons.ac.be

[3] UMons, Belgium, tarek.belabed@umons.ac.be

[4] senior IEEE, UMons, Belgium, carlos.valderrama@umons.ac.be

## 1 Abstract

*We need open platforms driven by specialists, in which queries can be created and collected for long periods and the diagnosis made, based on a rigorous clinical follow-up. In this work, we developed a multi-language robot interface helping to evaluate the mental health of seniors by interacting through questions. The specialist can propose questions, as well as to receive users' answers, in text form. The robot can automatically interact with the user using the appropriate language. It can process the answers and under the guidance of a specialist, questions and answers can be oriented towards the desired therapy direction. The prototype, was implemented on an embedded device meant for edge computing, thus it is able to filter environmental noise and can be placed anywhere at home. The experience is now available for specialists to create queries and answers through a Web-based interface.*

## 1. Introduction

Mental health care and diagnosis are today migrating towards mobile solutions [1][1][2][1]. Mobile applications provide a more accessible support [3]. This becomes particularly interesting, knowing that people dealing with mood, stress, or anxiety not always seek professional help or get care when it is really needed[4]. On the other hand, care or help is not always available when needed, for reasons such as location, financial averages or for societal reasons [5]. Examples of mobile applications are MIMOSYS [6] and CHADmon [7]. MIMOSYS [6] is a smartphone app that monitors mental health from the human voice detection diseases or disorders from emotional changes. The authors in [7] present CHADMon, a dedicated mobile application for voice analysis and monitoring of mental state and phase change detection.

The interest and techniques were already under study considering multiple aspects, from acceptability to clinical efficacy, through targeted therapies and clinical benefits [2]. Regarding applications, careful must be taken with diagnosis, that can be harmful and stigmatizing without specialized intervention [1]. Moreover, evaluation and experimental testing mechanisms are fundamental for a clinical and appropriate validation [8]. Indeed, we need open platforms driven by specialists, in which queries can be created and collected for long periods and the diagnosis made, based on a rigorous clinical follow-up.

---
[1]

Today, smartphones are popular and available for private usage. Some applications are attractive to users for the same reasons, in particular young adults and user looking for self-help support. However, that is not always the case with seniors not so familiar with the technology but still interested for free hands interaction such as a robot or voice.

Voice-enabled technologies are leading multiple domains, from automotive to home automation [9]. According to [10], 50 percent of searches based on voice by 2020, idem for smart speakers by 2022 [11]. Voice search tends to be more mobile and locally targeted because it is integrated with many mobile apps and devices. Many digital assistants are integrated with products that are part of our everyday life [10]. Microsoft integrated Cortana into Windows 10 for text and voice search. Amazon's Echo is ready to answer questions as well as to control other home devices. Voice assistants such as Amazon, Google Home or Sonos One are free; a number of requests are product searches that also offer placements to advertisers. Although these commercial products are not always open to customization, they are supported by development platforms, as in the case of Amazon AWS [12], Google [13] and IBM Watson [14], for example.

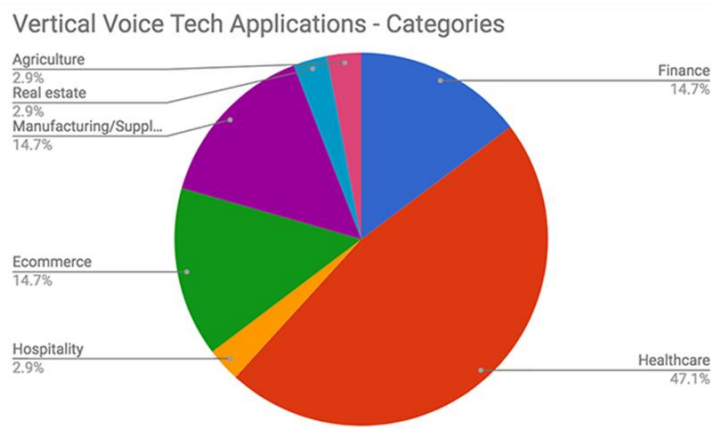

*Figure 1: Vertical Applications in Voice Tech* [15].

According to [15], the healthcare sector is the most popular category (47.1%) (**Figure 1**) for vertical voice-based applications. Telemedicine encourages conversation applications in the health field, particularly where hospitals have a strong incentive to provide high quality follow-up care. However, restrictions such as the confidentiality of the data involved and low error tolerance make it difficult to grow quickly. Thus, the high cost of physicians and caregivers is spent on hours of data collection in electronic health records. The voice health sector also extends to seniors who wish to stay at home, especially those who refuse mobile or smart technologies requiring dexterity or good vision [9]. Aging at home implies socializing, AI-based activity oriented interfaces and daily monitoring services.  Robot-based, patient-caregiver, communication saves time and therefore increases the productivity of already planned tasks such as remainders and appointments. Physician notes, such as the Electronic Health Record (EHR) and patient feedback, now use voice technology and AI-based natural language scribes [16] on multiple platforms (PC, smartphones), including new microphones and wearable voice interfaces [17].

This work followed several objectives: to provide a multilingual voice interaction platform, facilitating the specialist's intervention by creating protocols and text queries, as well as text forms for advice and collected results, to integrate and experiment existing technologies able to provide an automatic assessment of emotions providing graphical views of results' evolution, to pay attention to non-response situations, and to integrate the platform to a robot or a voice interface.

The article is organized as follows. Section 2 presents the state of the art in home health care voice-based products. We are particularly interested in embedded voice interfaces and devices, as well as the development tools available. Section 3 describes the system implemented. Section 4 presents the evaluation results. We conclude this document in section 5.

## 2. State of Art

Healthcare home products are evolving thanks to Ai-based platforms and on-line technologies (Figure 2). Several healthcare platforms are working with Amazon Alexa and Google Assistant smart speakers, for instance, Cuida Health LISA [18], a friendly voice assistant and companion who remembers medicines, appointments and monitors wellness status daily, RemindMeCare [19], Memory Lane [20] or Senter [21]. Some are AI-based social robots, such as ElliQ [22] and Senter, encouraging daily personalized activities. On-line technologies such as LifePod [23] enable schedules and voice services, providing valuable data to professionals and caregivers. Many are actually hands-free voice devices, such as Rosie Reminder [24] and ElliQ [22] or wearable, such as Notable [17]. AI-based natural language systems capture patient-physician interaction, prepare real-time patient notes in the exam room, and produce text-form EHRs [16] [17].

Voice assistants such as Amazon, Google or IBM Watson provide some libraries and APIs. These commercial products are not always open to customization but supported by development platforms. IBM Watson [14] proposes tools for speech (convert text and speech with the ability to customize models), language (analyze text and extract meta-data from unstructured content), empathy (understand tone, personality, and emotional state). As with Google, they also provide a *LanguageTranslator* for documents that can improved with the *NaturalLanguageClassifier*, a machine learning to analyze text and labels by organizing data into custom categories. The *ToneAnalyzer* library is intended to understand the emotions and style of communication in the text. The *PersonalytyInsights* library predicts personality characteristics, needs and values through written text.

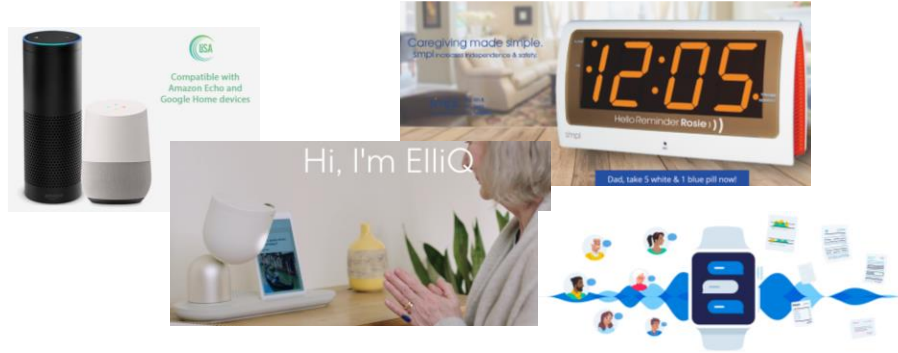

*Figure 2: Healthcare home products hands-free voice devices. Several healthcare platforms are working with Amazon Alexa and Google Assistant smart speakers, for instance, Cuida Health LISA [18] (left). Many are actually hands-free voice devices, such as Rosie Reminder [24] (right) and ElliQ [22] (center), others are wearable, as Notable [17] (bottom-right).*

## 3. Implementation architecture

To build a hands-free voice device, we target an edge computing embedded system (Figure 3). That system combines hardware libraries for audio processing and a Python program running on an ARM A9 CPU. The prototype was implemented on a Xilinx PYNQ-Z1 board [25], designed to be used as an open-source framework, enabling embedded programmers to exploit capabilities of reconfigurable hardware on the *APSoC* (*All programmable System-on-Chip*) Zynq family.

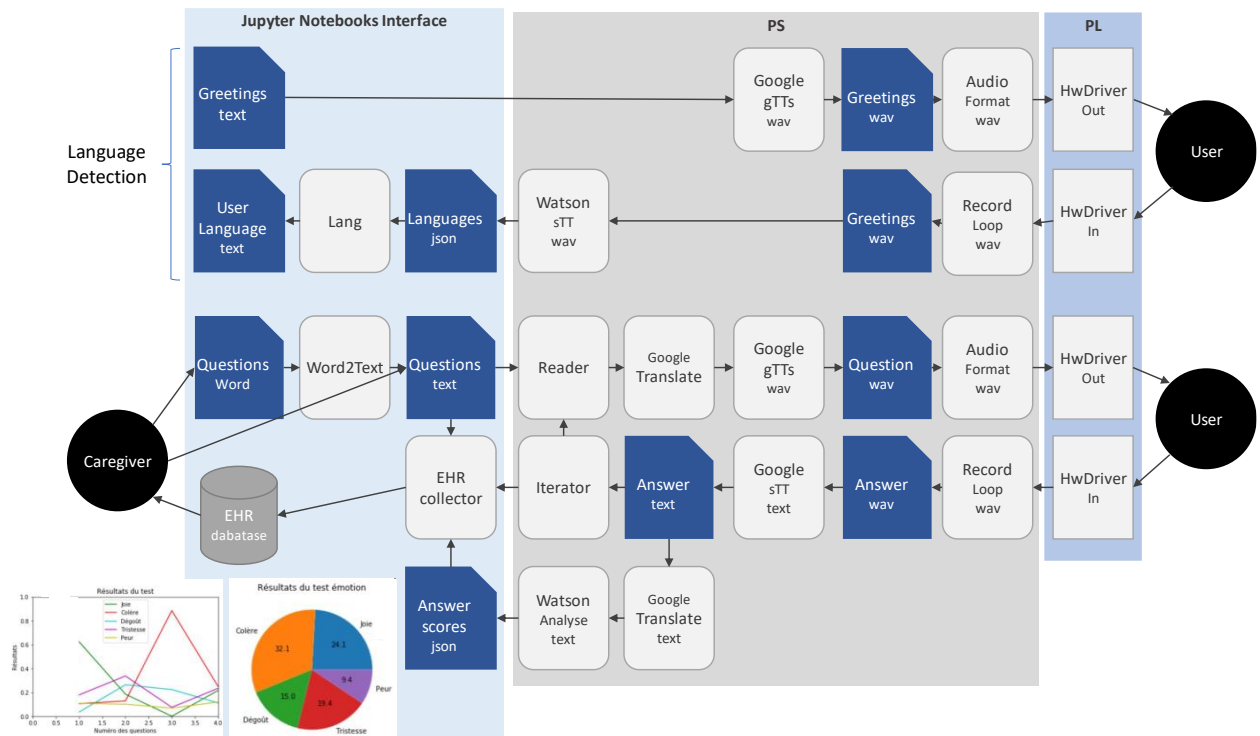

*Figure 3: Healthcare home hands-free voice device system architecture. The edge-computing embedded system is composed of three sections: Jupyter WEB interface, Programmable Software PS and Programmable Logic PL. The PL part uses real-time audio procession interfaces for user interaction using a headset. The Jupyter WEB interface allows to recover results in text*

*and graphical form. The first task of the system is to recognize the user language. The second steps to process the questions list that is used to fill the HER collector and database with the user answers.*

The software part (*Programmable Software PS*) of the APSoC is programmed using Python libraries from Google Cloud [13] (translate) and IBM Watson [14] (speech, empathy and natural language), all in a Jupyter Notebook [26] development environment. The hardware part (*Programmable Logic PL*) is a real-time audio processing programmable logic circuit, imported as a hardware library and programmed through an API, the same way as software. The platform can be accessed through a Web server hosting the Jupyter Notebooks design environment that includes the IPython kernel and packages running on a Linux OS.

### 3.1. Record a complete user response

The hardware API *Pynq.Record* is used to record the microphone input into an audio file. The audio driver (*HwDriver* on Figure 3) continually generates audio, but the API can only record a time interval. For this reason, the Python program continuously records 4 seconds each time, until there is no more incoming data (*Record Loop* on Figure 3). In this manner, we can record a complete answer to be send to *Google.SpeechToText*. This last operation can be intertwined, so we can build the response in text format during audio recording. We can also provide a playback of the answer by using the audio output and the recorded file.

### 3.2. Queries and Answers: audio and text formats

Queries or answers to the user are entered by the professional as text. This is a shortcut for the professional, he can write it using his own language and be translated according to the user. He can also use a Word document, with each question ended with a question mark. In that case, the questions will be added to a list on a text-format file. We use *Google.TextToSpeech*, to create the equivalent *mp3* audio file for each question. Alternatively, the professional can use the microphone to prepare his sets. The resulting audio must be converted to a *wav* file and adjusted to the driver parameters (24-bit, 48-kHz, 2-channel) using *Subproces* (*Audio Format* on Figure 3). One option was to use Audacity [24], but then, it had to be integrated in the Python program and follow 2 conversion steps: from mono to stereo, then to 24 bits. Anyway, it was used to validate the results of the conversion. The same can be realized by using *PyPI.PyDub.AudioSegment* [27] conversion from mono to stereo and *PyPI.SoundFile* [28] conversion from 16 bits 44 KHz to 24 bits 48 KHz.

### 3.3. Text from user responses in the appropriate language: language detection

Based on queries, we can collect user answers through the audio interface. It can be just recorded as an audio file. Instead of simply creating a multilingual multimedia HER, another solution is to obtain user responses in text form and in an appropriate language. This format facilitates the search for keywords and features. By using *Google.SpeechToText.recognize_google*, along with the audio file and language as input parameters we can get the text form. The file can then be translated into the language desired by the specialist using *Google.Translator.translate* for further processing of text-form recordings. In a similar manner, we can recover an audio file. This is especially important when using tools only available for English input, as we will show later. The library *Google.TextToSpeech.gTTs* will produce a *wav* format audio file at the desired speech rate.

### 3.4. Artificial intelligence and Emotions on the spot

The IBM Watson [14] proposed libraries for speech, language and empathy are based on Ai and machine learning engines. Some libraries are available in Python, however, most have paid access. So, we limited the experience to just language translation and tone analysis. As with Google, they also provide a *LanguageTranslator*. The *ToneAnalyzer* library is intended to understand the emotions and style of communication in the text. The *Analyze* library processes a text document based on *emotion* and *sense* parameters to focus on and provides a *json* answer with a score of confidence according to 6 alternative results: joy, anger, disgust, fear, sadness and positivity/negativity. In this case, the experiment consisted of detecting the language used, converting any information into that language, initiating the audio exchange between the specialist (questions and advice) and the user (answers) and finally to provide an emotional score.

### 4. Implementation results

The full system was implemented on a PC and on the Pynq board for evaluation purposes. The prototype uses a Jupyter Notebook Web interface for the full process, in this manner, we can make changes during the development process. However, the final version just provides the tools necessary for the specialist to enter queries and advice, receive user responses and graphical results.

The language detection (*Language Detection* on Figure 3) consists of a welcome text sentence that is transcribed into audio using *Google.TextToSpeech*, then we detect the language of the user's answer with IBM Watson *SpeechToText* that returns a json file containing several evaluated languages, the highest scored language is selected for the rest of the process.

The answers collection process (*Iterator* on Figure 3) was implemented on a PC and the Pynq board for evaluation purposes. The process consists in processing the list of questions (each question is transformed and sent to audio output) and user answers (each answer received on the device audio input is transformed to a text-format). The process detects when the user does not answer a question, in that case, the question is repeated, otherwise we continue with the next question until the end of the list.

We are also interesting on the results provided by IBM Watson, for that reason we translate the answer to English (the only language accepted by Watson) and send it to the Analyzer, who returns a *json* answer with a score of confidence according to 6 alternative results (*Watson Analyze* on Figure 3). We attach the score to each answer and calculate an average score for the set. Finally, we provide multimedia HER with audio and text responses, as well as a graphical view of the result obtained with IBM Watson. We provide a show a pie chart that shows the means of emotions calculated and a graph that shows us all the percentages collected for each question.

We performed some evaluation tests considering that the platform implemented on a PYNQ board is less powerful than a PC. We use *SpeechRecognition* (pip3 install speech-recognition) [29], *Google Text to Speech gTTs* (pip3 install gTTs)[29], *json* and IBM Watson *Natural Language Understanding* (pip3 install NaturalLanguageUnderstandingV1)[30] libraries for information processing. On the PC we also used *TempFile* (pip3 install tempfile)[31] and *PyGame* (pip3 install pygame) for audio files processing. On Pynq we use *SoundFile* (pip3 install soundfile) [32] and *PyDub* (pip3 install pydub)[33] to manipulate audio files. In addition, we use *Time* to create pauses during execution and *Numpy* for the graphics. We first evaluate the processing speed compared to an I5 processor. The results show that, in the worst case, the board spends 21 seconds per question compared to the 37 seconds on the PC.

| Files | % recognition | %joy | % anger | % sadness | % fear | % disgust | emotion expected | final emotion |
|---|---|---|---|---|---|---|---|---|
| "heureux.wav" | 100 | 0,87 | 0,01 | 0,04 | 0,01 | 0,01 | joy | JOY |
| "malheureuse.wav" | 100 | 0,09 | 0,05 | 0,72 | 0,07 | 0,06 | sadness | SADNESS |
| "colere.wav" | 100 | 0,02 | 0,85 | 0,04 | 0,02 | 0,02 | anger | ANGER |

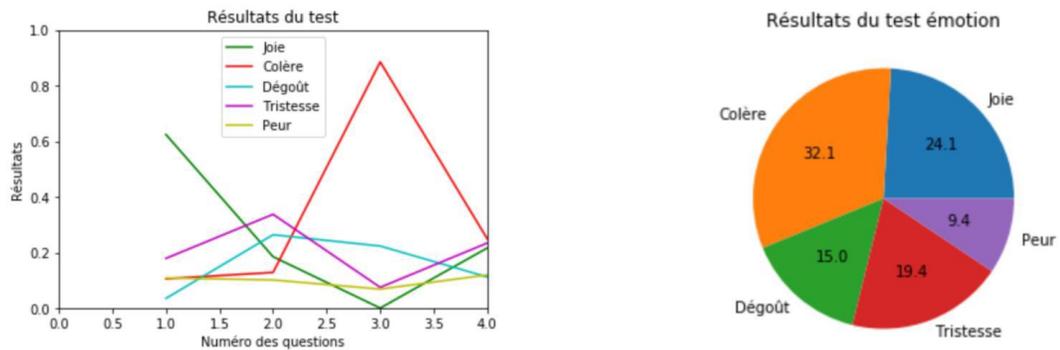

*Figure 4: Test including the IBM Watson analyzer. The user answers in French language followed the full process including the Watson Analyse API. The table shows some results. The Jupyter Notebooks WEB interface provides as well a graphical view: a pie chart shows the means of emotions and a graph with all the percentages collected for each question.*

We were also interested on the utilization of IBM Watson. For that reason, we created some user answers in wav audio format to see the results that analyze can provide. The test was no exhaustive but at least helped us to understand what we expect from that tool. We used three audio files: heureux.wav *"I'm so happy to live here"*, malheureuse.wav *"I hate this world"* and colère.wav *"I can't tolerate this. I don't understand why people do that. Certainly, the content was originally in French language, but after the full processing the results were satisfactory. Other tests were performed with results similar to the preliminary test shown on* Figure 4*.*

5. **Conclusion**

In this work, we developed a multi-language robot interface helping to evaluate the mental health of seniors by interacting through questions. The prototype, implemented on an embedded device is meant for edge computing. The platform is able to process text form queries from the caregiver and collect user answers. The device can also filter environmental noise and be placed anywhere at home. The experience is now available for specialists to create queries and answers through a Web-based interface. Queries can be created and collected for long periods and the diagnosis made, based on a rigorous clinical follow-up. The specialist can propose questions, as well as to receive users' answers, in text form. The robot can automatically interact with the user using the appropriate language. It can process the answers and under the guidance of a specialist, questions and answers can be oriented towards the desired therapy direction.

## 3   Acknowledgments

The authors would also like to acknowledge the contribution of the COST Action CA16226 Indoor living space improvement: Smart Habitat for the Elderly.